\documentstyle[12pt]{article}

\newcommand{\be}{\begin{equation}}
\newcommand{\ee}{\end{equation}}

\def\gsim{\:\raisebox{-0.5ex}{$\stackrel{\textstyle>}{\sim}$}\:}

\begin{document}  
\renewcommand{\thefootnote}{\fnsymbol{footnote}}

\pagestyle{plain}
\begin{flushright}
TUM-HEP-369-00 \\
March 2000
\end{flushright}

\vspace{1cm}

\begin{center}

{\large \bf Comment on ``A New Dark Matter Candidate: Non--thermal
Sterile Neutrinos''}

\vspace{7mm}
Manuel Drees \\

\vspace*{5mm}

{\it Physik Dept., TU M\"unchen, James Franck Str., D--85748
Garching, Germany}
\end{center}

\vspace{2cm}

\begin{abstract}

\noindent 
I point out that the sterile neutrinos suggested as candidates for
``cool'' Dark Matter will decay through their mixing with light
neutrinos. This leads to an upper bound of about 200 keV on the mass
of the sterile neutrinos, but might facilitate their detection.

\end{abstract}

\vspace{2cm}

Recently Shi and Fuller proposed \cite{1} a new kind of Dark Matter
candidate: a sterile neutrino, with small but non--vanishing mixing
with the ordinary neutrinos. The idea is that a large pre--existing
lepton asymmetry stored in light neutrinos can be converted into an
asymmetry of massive sterile neutrino through a thermal
resonance. Since low--energy neutrinos would be converted first, the
resulting relic neutrinos are less energetic (``cooler'') than thermal
relics of the same mass would be. This mechanism requires \cite{1} a
lepton asymmetry $\Delta L \gsim 10^{-3}$ and a mixing angle between
active and sterile neutrinos $\sin^2 2 \theta \gsim 10^{-9}$. One then
obtains the desired relic density for $m_{\nu_s} \sim 1 \ {\rm keV}
\cdot ( 0.1 / \Delta L)$. An explicit particle physics model along
these lines has been constructed by Chun and Kim \cite{2}. Here the
sterile ``neutrino'' is the axino, which mixes with ordinary neutrinos
through R--parity violating interactions.

These studies seem to have overlooked the simple fact that the mixing
with ordinary, light neutrinos allows the sterile neutrino to decay
into three light neutrinos through the exchange of a virtual $Z$
boson. The total decay width of the sterile neutrino is then found to
be
\be \label{e1}
\Gamma(\nu_s \rightarrow 3 \nu) = \Gamma(\mu^- \rightarrow e^- \nu_\mu
\bar{\nu}_e ) \cdot \sin^2 \theta_{\rm eff} \cdot \left( \frac
{m_{\nu_s}} {m_\mu} \right)^5,
\ee
where $\sin^2 \theta_{\rm eff} = \sum_{i=1}^3 \sin^2 \theta_{\nu_s
\nu_i}$. When calculating this decay width one has to keep in mind
that two diagrams contribute to the decay into three neutrinos of
equal flavor. The corresponding matrix elements are in fact identical
to each other, leading to a factor of 2 enhancement of $\Gamma(\nu_s
\rightarrow \nu_i \nu_i \bar{\nu}_i)$ relative to $\Gamma(\nu_s
\rightarrow \nu_i \nu_j \bar{\nu}_j), \ i \neq j$.

From eq.(\ref{e1}) one computes the lifetime
\be \label{e2} 
\tau_{\nu_s} = 8.5 \cdot 10^6 \ {\rm yrs} \cdot \frac
{1} {\sin^2 \theta_{\rm eff} } \cdot \left( \frac {10 \ {\rm keV}}
{m_{\nu_s}} \right)^5.
\ee
Requiring this lifetime to exceed $2 \cdot 10^{10}$ yrs then leads to
the bound
\be \label{e3}
\sin^2 \theta_{\rm eff} < 4.3 \cdot 10^{-4} \left( \frac {10 \ {\rm
keV}} {m_{\nu_s}} \right)^5,
\ee
which limits the allowed parameter space of this kind of model. In
particular, the lower bound on the mixing angle, which follows from
the requirement of adiabatic $\nu_i \rightarrow \nu_s$ conversion,
$\sin^2 \theta_{\rm eff} > 10^{-10}$, implies
\be \label{e4}
m_{\nu_s} \leq 200 \ {\rm keV}.
\ee
This is well above the range of masses considered in
refs.\cite{1,2}. However, there higher masses were not considered for
a purely technical reason: if $m_{\nu_s} \gsim 20$ keV, the resonance
conversion temperature exceeds the temperature for the QCD phase
transition, making the calculation more complicated. The bounds
(\ref{e3}) and (\ref{e4}) are independent of this consideration. [In
the model of ref.\cite{2} the axino can also decay into a gravitino,
but the corresponding partial width is sufficiently small \cite{2}.]

The decay of the sterile neutrino into three light, active neutrinos
not only leads to a new constraint on the model parameters, it also
might allow to detect this kind of ``cool'' Dark Matter through its
decay products. The resulting flux of active neutrinos can be
estimated as
\be \label{e5}
\Phi(\nu_i) \sim \frac {10^{10} \ {\rm yrs}} {\tau_{\nu_s}} \cdot \frac
{10 \ {\rm keV}} {m_{\nu_s}} \cdot \Omega_{\nu_s} \cdot \frac
{10^{11}} { {\rm cm}^2 \, {\rm sec}}.
\ee
where $\Omega_{\nu_s}$ is the $\nu_s$ relic density in units of the
critical density. Detecting this flux will be very difficult, given
the low energy of the neutrinos ($E< m_{\nu_s}$). Note also that the
flux (\ref{e5}) is smaller than the flux of solar neutrinos, unless
$\tau_{\nu_s}$ is very close to its lower bound. Nevertheless
detecting the decay products of $\nu_s$ does not appear to be quite as
hopeless as the direct detection of these sterile relics.

\bigskip

\noindent {\bf Acknowledgements:} \\
This work was supported in part by the ``Sonderforschungsbereich 375--95 f\"ur
Astro--Teilchenphysik'' der Deutschen Forschungsgemeinschaft.

\end{document}